\documentclass[a4paper,11pt]{article}
\usepackage[english]{babel}
\usepackage{amsfonts,slashed}
\usepackage{url}
\usepackage{latexsym}
\usepackage{amsmath}
\usepackage{amssymb}
\usepackage{indentfirst}
\usepackage{graphicx}
\usepackage{epsfig}
\usepackage{amsthm}
\usepackage{amsfonts}
\usepackage{indentfirst}
\usepackage[colorlinks]{hyperref}
\usepackage{cite}
\usepackage{cancel}
\usepackage{tikz}
\usepackage{enumitem}
\usepackage{float}

\usepackage{upgreek}

\begin{document}

\begin{titlepage}

\title{\bf{On SIR-type epidemiological models and population heterogeneity effects}} 
\author{Silke Klemm$^{1,2}$\thanks{silke.klemm@mi.infn.it} \ and Lucrezia Ravera$^{3,4}$\thanks{lucrezia.ravera@polito.it} \\ \\
{\small $^{1}$\textit{Dipartimento di Fisica, Universit\`{a} di Milano, Via Celoria 16, 20133 Milano, Italy}}\\
{\small $^{2}$\textit{INFN, Sezione di Milano, Via Celoria 16, 20133 Milano, Italy}}\\
{\small $^{3}$\textit{DISAT, Politecnico di Torino, Corso Duca degli Abruzzi 24, 10129 Torino, Italy}}\\
{\small $^{4}$\textit{INFN, Sezione di Torino, Via P. Giuria 1, 10125 Torino, Italy}}}
\clearpage\maketitle
\thispagestyle{empty}

\begin{abstract}

In this paper we elaborate on homogeneous and heterogeneous SIR-type epidemiological models. We
find an unexpected correspondence between the epidemic trajectory of a transmissible disease in a 
homogeneous SIR-type model and radial null geodesics in the Schwarzschild spacetime.
We also discuss modeling of population heterogeneity effects by considering both a one- and
two-parameter gamma-distributed function for the initial susceptibility distribution, and deriving the 
associated herd immunity threshold. We furthermore describe how mitigation measures can be taken into 
account by model fitting.

\end{abstract}

\vspace{2cm}


\end{titlepage}

\tableofcontents

\noindent\hrulefill

\section{Introduction}

The SARS-CoV-2 pandemic led, in many countries, to lockdown measures aiming to control and limit
the spreading of the virus. A key role when facing global events of this type is played by mathematical modeling of infectious diseases, which allows direct validation with real data. This consequently permits to evaluate the effectiveness of control and prevention strategies, giving support to public health.

In this context, Susceptible-Infected-Removed (SIR) models of epidemics (see e.g. 
\cite{Kermack:1927xxx,Mickens:2012xxx}) capture key features of a spreading epidemic
as a mean field theory based on pair-wise interactions between infected and susceptible individuals,
without aiming to describe specific details. In particular, in the presence of $I$ infected individuals in a population of $N$ individuals, the infection can be transmitted to susceptible individuals $S$. They stay infectious during an average time $\gamma^{-1}$, after which they no longer contribute to infections.
The fraction of immune individuals in the population beyond which the epidemic can no longer grow defines the herd immunity threshold (HIT).

Simple SIR models commonly assume the population to be homogeneous; each individual has the same probability of being infected by the disease. However, in order to take into account that the
infection probability actually depends on age, sex, connections with other individuals, etc., SIR models for 
heterogeneous populations have been considered \cite{Gomes:2020xxx,Aguas:2020xxx,
Montalban:2020xxx,Neipel:2020xxx}. In these models, a parameter, usually denoted by $\alpha$, is commonly introduced to describe population heterogeneity and, hence, variation in susceptibility of individuals.

Studying the transmission of the virus SARS-CoV-2, in \cite{Gomes:2020xxx} it was shown that the percentage of a homogeneous population to be immune given some value for $R_0$ (which is the basic reproduction number, namely the average number of new infected generated by an infected individual at the early epidemic stage) noticeably drops if the population is considered to be highly heterogeneous. More specifically, while herd immunity is expected to require 60-75 percent of a homogeneous population to be immune given an $R_0$ (that is the basic reproduction number) between 2.5 and 4, these percentages drop to the 10-20 percent range for the coefficients of variation in susceptibility considered in \cite{Gomes:2020xxx} between 2 and 4. In particular, it was shown that individual variation in susceptibility or exposure (connectivity)
accelerates the acquisition of immunity in populations due to selection by the force of infection.
More susceptible and more connected individuals have a higher propensity to be infected and thus are likely to become immune earlier. Due to this selective immunization, heterogeneous populations require less infections to cross their HITs than homogeneous (or not sufficiently heterogeneous) models would suggest.
In \cite{Gomes:2020xxx} the initial susceptibility was considered to be gamma-distributed, with a one-parameter gamma distribution. Besides, the case of a lognormal distribution was treated numerically. The gamma distribution was also considered in \cite{Duchesne:2022xxx} to model the first-wave COVID-19 daily cases, and it was proven, in this context, to provide better results than the Gaussian, Weibull (and Gumbel) distributions.

Taking into account heterogeneity effects has proven to be relevant also in the spread of smallpox (cf. \cite{Neipel:2020xxx}), where homogeneous models are not capable to explain the data, as well as for tuberculosis and malaria (see, e.g., \cite{Gomes:2020xxx} and references therein).

In this work we discuss modeling of population heterogeneity effects by considering both a one-parameter gamma-distributed function and a two-parameter one for the initial susceptibility distribution, deriving the associated HIT. The latter is computed analytically in both cases. We also describe a possible way to take into account mitigation measures when performing model fitting in the case of the one-parameter initial gamma distribution, while the two-parameter initial gamma distribution appears to automatically accommodate this external action on diseases spread. 
On the other hand, regarding homogeneous SIR models, we present an intriguing feature of a simple model of this type, which paves the way to future analytically tractable studies of epidemiological models.

The remainder of this paper is structured as follows: In Section \ref{homsir}, we review homogeneous SIR-type models of epidemics and, in Section \ref{nullgeod}, we present a correspondence between the epidemic trajectory in a homogeneous SIR model and radial null geodesics in the Schwarzschild spacetime. Subsequently, in Section \ref{hetsir}, we discuss modeling of population heterogeneity effects to capture the fact that the probability of being infected is not the same for all individuals. Section \ref{final} is devoted to final remarks and possible future developments of our analysis.

\section{Homogeneous SIR-type models}\label{homsir}

In an SIR-type model \cite{Kermack:1927xxx}, the population is divided into susceptible, infected and 
recovered individuals, whose numbers are denoted respectively by $S$, $I$, and $R$. Their dynamics
is governed by the equations
\begin{equation}
\dot S = - f(I,S)\,, \qquad\dot I = f(I,S) - g(I)\,, \qquad\dot R = g(I)\,. \label{SIR-gen}
\end{equation}
Here $f(I,S)$ denotes the infection force, i.e., the rate at which susceptible persons acquire the infectious 
disease, while $g(I)$ is some function to be specified
below. The upper dot symbol denotes the time derivative. From \eqref{SIR-gen} one obtains the conservation law
\begin{equation}
\dot S + \dot I + \dot R = 0\quad\Rightarrow\quad S + I + R = \text{const.} = N\,,
\end{equation}
with $N$ the total number of individuals in the population. A common choice is $f(I,S)=\beta I S$,
$g(I)=\gamma I$, where $\beta$ is the transmission (or infection) rate (per capita),\footnote{The infection rate $\beta$ can in general depend on time $t$; this time dependence could correspond to seasonal changes or mitigation measures \cite{Dehning:2020xxx,Flaxman:2020xxx,Kissler:2020xxx}.} and $\gamma$
denotes the rate of recovery. It is related to the average recovery time $D$ by $D=1/\gamma$.
We have thus\footnote{In this work, we multiply the quantity $\beta$ with the constant factor $N$ with respect to the one defined, e.g., in \cite{Neipel:2020xxx}, that is $\beta \to N \beta$.}
\begin{equation}\label{SIR-common}
\dot S = -\beta IS\,, \qquad\dot I = \beta IS - \gamma I\,, \qquad\dot R = \gamma I\,.
\end{equation}
This implies
\begin{equation}
\frac{dI}{dS} = -1 + \frac{\gamma}{\beta S}\,, \label{dIdS}
\end{equation}
which can be integrated to give the epidemic trajectory
\begin{equation}
I - I_0 = S_0 - S + \frac{\gamma}{\beta}\ln\frac S{S_0}\,, \label{epid-traj}
\end{equation}
with $I_0=I(t=0)$ and $S_0=S(t=0)$. In order to obtain the early growth of the epidemic, one linearizes \eqref{SIR-common} around $S=S_0\approx N$ and $I\approx 0$, i.e., sets
\begin{equation}
S = N - \delta S\,,\qquad I = \delta I\,,\qquad\qquad\delta S,\delta I\ll N\,. \label{linearized}
\end{equation}
This leads to the exponential law
\begin{equation}
\delta I = I_0\,e^{\gamma (R_0-1) t}\,, \label{deltaI-t}
\end{equation}
where 
\begin{equation}
    R_0=\frac{N\beta}{\gamma}
\end{equation}
is the basic reproduction number. It denotes the average number of new
infections generated by an infected individual (at the early epidemic stage).

\begin{figure}[t]
    \centering
    \includegraphics[scale=1]{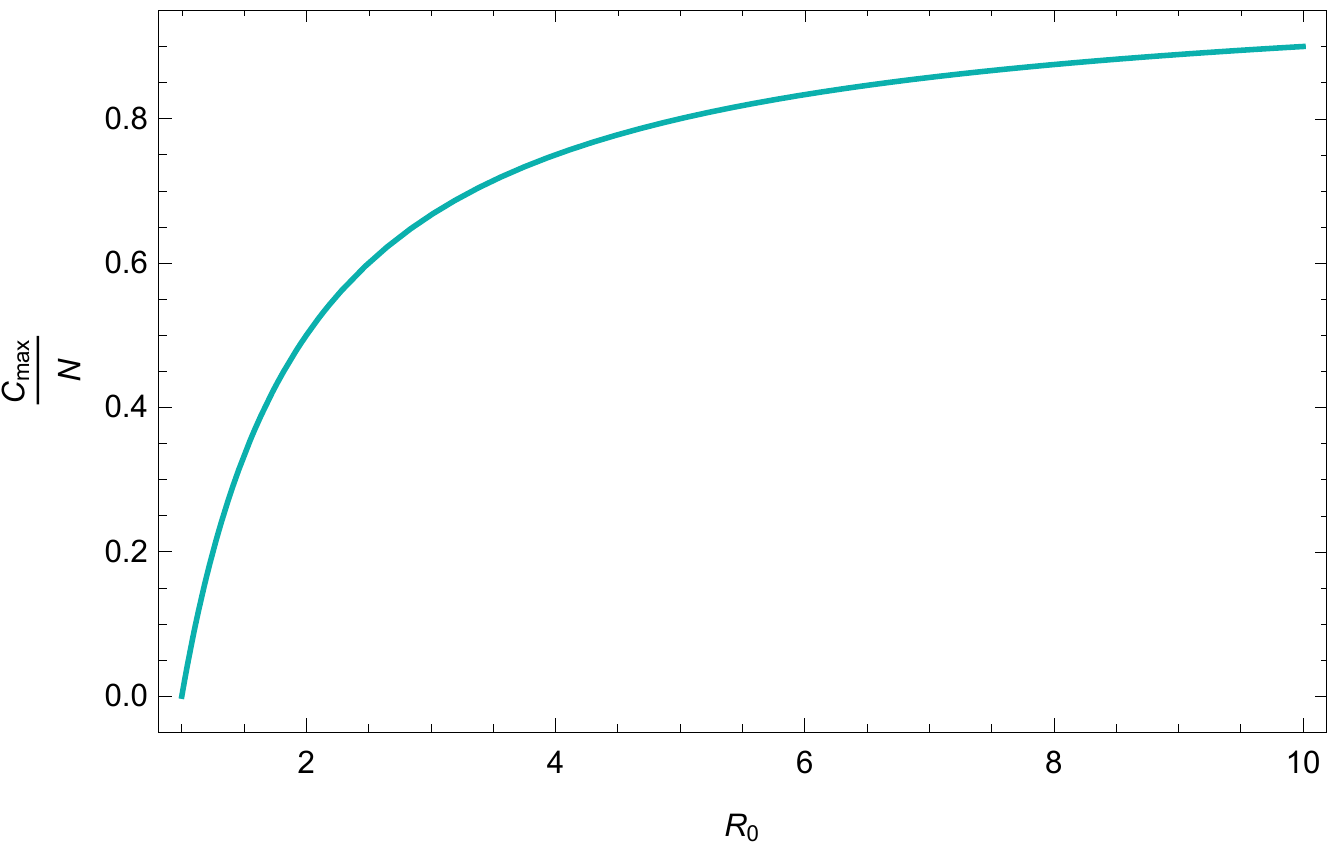}
    \caption{Herd immunity threshold $\frac{C_{\text{max}}}{N}$ of a homogeneous population as a function of $R_0$ (more specifically, here we consider $1\leq R_0 \leq 10$). The larger $R_0$, the more the HIT rises.}
    \label{fig:hvR0}
\end{figure}

The function $I(S)$ has a maximum at $S=S_{\text{max}}=\gamma/\beta$.
At this peak, a fraction $S_{\text{max}}/N = 1/R_0$ of individuals remains susceptible. The cumulative
number of infections $C=N-S$ at the  maximum of $I$ thus obeys
\begin{equation}
\frac{C_{\text{max}}}N = 1 - \frac1{R_0}\,. \label{herd-imm-hom}
\end{equation}
This is the well-known formula for the herd immunity level (or herd immunity threshold, HIT), i.e., the fraction of immune individuals in the population beyond which the epidemic can no longer grow. 
Here we are not considering mitigation measures, nor reinfections. Hence, in particular, the threshold to reach herd immunity is estimated by considering natural infections without restrictions (lockdown, social distancing, etc.) and without taking into account possible vaccinations.
The plot in Figure \ref{fig:hvR0} displays the herd immunity level \eqref{herd-imm-hom} as a function of $R_0$.

When the epidemic stops we have $I=0$. Using \eqref{epid-traj},
it is easy to show that the number of susceptible individuals left over at the end of an epidemic is given by
\begin{equation}
S = -\frac{S_0}{R_{\text e}} W_0\left[-R_{\text e}\exp\left(-R_{\text e}\left(1 +\frac{I_0}{S_0}\right)\right)
\right]\,, \label{S-final}
\end{equation}
where
\begin{equation}
    R_{\text e}=S_0\beta/\gamma =S_0 R_0/N \label{ern}
\end{equation}
is the effective reproduction number, while $W_0$ represents a 
particular branch of the Lambert W-function, which is defined as the inverse of the function
$f:x\mapsto x e^x$.

Let us also mention that a different choice for the functions $f$ and $g$ in \eqref{SIR-gen} was made by
Mickens\cite{Mickens:2012xxx}, namely $f=\beta\sqrt{IS}$ and $g=\gamma\sqrt I$, and thus
\begin{equation}
\dot S = -\beta\sqrt{IS}\,, \qquad\dot I = \beta\sqrt{IS} - \gamma\sqrt I\,, \qquad\dot R =
\gamma\sqrt I\,. \label{SIR-Mickens}
\end{equation}
The square root leads to a power-law (instead of exponential) early growth of the epidemic. Indeed,
linearizing \eqref{SIR-Mickens} according to \eqref{linearized}, one gets
\begin{equation}
\delta I = \left((\beta\sqrt N  - \gamma)\frac t2 + I_0\right)^2\,.
\end{equation}
The model \eqref{SIR-Mickens} is therefore particularly adapted to describe effects of mitigation
measures imposed by governments, like social distancing and so on. 
From \eqref{SIR-Mickens} we obtain
\begin{equation}
\frac{dI}{dS} = -1 + \frac{\gamma}{\beta\sqrt S}\,,
\end{equation}
which leads to the epidemic trajectory
\begin{equation}
I = I_0 - (S - S_0) + \frac{2\gamma}{\beta}(\sqrt S - \sqrt{S_0})\,.
\end{equation}
This has a maximum at $S=S_{\text{max}}=(\gamma/\beta)^2$. In this case, the formula for the herd immunity level becomes
\begin{equation}
\frac{C_{\text{max}}}N = 1 - \frac{\gamma^2}{\beta^2 N}\,.
\end{equation}
Observe that the latter depends, in particular, on the squared of the ratio $\gamma/\beta$, and therefore may be also rewritten as $\frac{C_{\text{max}}}N = 1 - \frac{N}{R_0^2}$.

\subsection{Epidemic trajectory as radial null geodesics in Schwarzschild spacetime}\label{nullgeod}

Remarkably, the epidemic trajectory \eqref{dIdS} coincides with radial null geodesics in the Schwarzschild 
spacetime,\footnote{The Schwarzschild metric describes a static, non-rotating black hole solution to the Einstein's field equations. It was discovered by Karl Schwarzschild within a year of Einstein’s publication of the theory of general relativity. A null geodesic is the path that a massless particle, such as a photon, follows. It is called null since its interval (its ``distance'' in four-dimensional spacetime) is equal to zero and it does not have a proper time associated with it.} which are given by
\begin{equation}
0 = -\left(1 - \frac{2m}r\right) d\uptau^2 + \frac{dr^2}{1 - \frac{2m}r}\quad\Rightarrow\quad
d\uptau = \pm\frac{dr}{1 - \frac{2m}r}\,, \label{null-geod}
\end{equation}
where $m$ is the black hole mass, $r$ the radial coordinate, and $\uptau$ the time coordinate (time measured by a stationary clock at infinity).\footnote{We adopt geometrized units, that is $c=G=1$, where $c$ is the speed of light in vacuum and $G$ the gravitational constant.} If we identify
\begin{equation}
I = \mp a \uptau \,, \qquad S = a(r - 2m)\,, \label{IStaur}
\end{equation}
with $a$ an arbitrary scale factor, \eqref{dIdS} becomes precisely \eqref{null-geod}, provided that
$-2ma=\gamma/\beta$. One can also map the evolution equations \eqref{SIR-common} into
\begin{equation}
\frac{d\uptau}{d\lambda} = \frac E{1 - \frac{2m}r}\,,\qquad\frac{dr}{d\lambda} = \pm E\,, \label{geod-lambda}
\end{equation}
that follow respectively from the conservation law $g_{\mu\nu}u^\mu\xi^\nu=-E$ and
$g_{\mu\nu}u^\mu u^\nu=0$. Here $\lambda$ denotes an affine parameter,
$E$ is the conserved energy, $u=d/d\lambda$ is tangent to the geodesic, while $\xi=\partial_\uptau$ 
is the timelike Killing vector. Using \eqref{IStaur}, the eqns.~\eqref{geod-lambda} reduce to
\eqref{SIR-common} if $E=\mp 1/a$ and
\begin{equation}
\frac{d\lambda}{dt} = \beta I S\,. \label{def-lambda}
\end{equation}
The scale factor $a$ is thus related to the energy $E$ of the geodesic, and $t$ is not an affine parameter.
Since $\beta IS$ is the infection force in this specific model, we have from \eqref{def-lambda}
that $\dot\lambda=f(I,S)$, so that $\lambda$ can be interpreted as infection momentum.
Moreover, from the first of \eqref{SIR-common} one gets $\dot S + \dot\lambda=0$, and therefore
$S+\lambda$ is constant. If we choose this constant to be equal to $N$, then
\begin{equation}
\lambda = N - S\,,
\end{equation}
which is the cumulative number of infections $C$. 

This unveiled correspondence between the epidemic trajectory \eqref{dIdS} and radial null geodesics in the Schwarzschild geometry may be useful to obtain new analytically tractable epidemiological models.

\section{Modeling population heterogeneity effects}\label{hetsir}

In order to capture the fact that the probability of being infected is not the same for all individuals,
we use the model of \cite{Neipel:2020xxx}, i.e., we introduce a distribution $s$ of susceptibilities $x$,
and denote by $s(x,t)dx$ the number of individuals with susceptibility between $x$ and $x+dx$
at time $t$. The total number of susceptible individuals reads
\begin{equation}
S(t) = \int_0^\infty s(x,t) dx\,, \label{S-int-s}
\end{equation}
and $s$ obeys
\begin{equation}
\frac{\partial s}{\partial t} = -\beta x I s\,, \label{dyn-s}
\end{equation}
which generalizes the first of \eqref{SIR-common}. Integrating \eqref{dyn-s} over $x$ leads to
\begin{equation}
\dot S(t) = -\beta\bar x(t) I S\,, \label{dot-S-barx}
\end{equation}
where the average susceptibility $\bar x(t)$ is given by\footnote{Notice that the average infection susceptibility $\bar x$, which is introduced to capture effects of population heterogeneity, also allows to modulate the infection rate $\beta$.}
\begin{equation}
\bar x(t) = \frac1{S(t)}\int_0^\infty s(x,t) xdx\,. \label{av-susc}
\end{equation}
In the special case $\bar x=1$, \eqref{dot-S-barx} boils down to the corresponding
equ.~in \eqref{SIR-common}. The dynamical equations are completed by
\begin{equation}
\dot I(t) = \beta\bar x(t) I S - \gamma I\,. \label{I(t)-hetero}
\end{equation}
The time course of an epidemic is often provided as the number of new cases per day. This
corresponds to the rate of new infections per unit time,
\begin{equation}
    J = \beta \bar{x} IS \,, \label{Jeq}
\end{equation}
with $J=-\dot{S}$.

Notice that the authors of \cite{Gomes:2020xxx,Aguas:2020xxx,Montalban:2020xxx} considered
two cases, namely variable susceptibility or variable connectivity (individuals that have many contacts
are both more likely to get infected and to infect others). They describe these situations with a
susceptible-exposed-infected-recovered (SEIR) model,
\begin{eqnarray}
\partial_t s(x,t) &=& -x\rho(t) s(x,t)\,, \qquad \partial_t e(x,t) = x\rho(t) s(x,t) - \delta e(x,t)\,,
\nonumber \\
\partial_t i(x,t) &=& \delta e(x,t) - \gamma i(x,t)\,, \qquad \partial_t r(x,t) = \gamma i(x,t)\,,
\label{SEIR-Gomes}
\end{eqnarray}
where
\begin{equation}
\rho(t) = \left\{\begin{array}{ll}\beta\int i(x,t) dx & \quad\text{variable susceptibility}, \\
\beta\int i(x,t) x dx & \quad\text{variable connectivity}, \end{array}\right.
\end{equation}
while $\gamma$ and $\delta$ are constants, the latter denoting the rate of progression from exposed
to infectious. For variable susceptibility, the first of \eqref{SEIR-Gomes}
is identical to \eqref{dyn-s}, if we set $I(t)=\int i(x,t) dx$. We see that in \eqref{SEIR-Gomes},
also $E(t)$, $I(t)$ and $R(t)$ are divided into infinitely many compartments $e(x,t)$, $i(x,t)$ and $r(x,t)$,
which is different form the model used in \cite{Neipel:2020xxx}, where only $S(t)$ is split.
In this paper, we shall limit ourselves to the case where only effects of heterogeneity in the degree of susceptibility to infection are taken into account, as it was done in \cite{Neipel:2020xxx}.

In what follows, it will prove useful to introduce a new time variable $\tau$, defined by \cite{Neipel:2020xxx}
\begin{equation}
    \dot\tau=\beta I \,. \label{dottaudef}
\end{equation}
Let us stress that $\tau=\tau(t)$. Equ. \eqref{dyn-s} can then
easily be integrated, and \eqref{S-int-s} gives
\begin{equation}
S(\tau) = \int_0^\infty s_0(x) e^{-\tau x} dx \,. \label{Laplace-transf}
\end{equation}
In other words, $S(\tau)$ is the Laplace transform of the initial distribution $s_0(x)$. 

\subsection{HIT for the case of a one-parameter gamma distribution}

As in \cite{Gomes:2020xxx,Aguas:2020xxx,Montalban:2020xxx,Neipel:2020xxx}, we shall now assume
that $s_0(x)$ is gamma-distributed with shape parameter $\alpha>0$ (here we start by considering, following \cite{Gomes:2020xxx,Aguas:2020xxx,Montalban:2020xxx,Neipel:2020xxx}, the case of a one-parameter, that is $\alpha$, gamma distribution with rate parameter $\eta:=\frac{1}{\theta}=\alpha$, being $\theta$ the scale parameter),
\begin{equation}
s_0(x)= S_0\frac{\alpha^\alpha}{\Gamma(\alpha)} x^{\alpha - 1} e^{-\alpha x}\,,
\end{equation}
where $S_0$ is the initial value of susceptible individuals and $\Gamma(\alpha)$ denotes Euler's gamma
function. This gives for the Laplace transform \eqref{Laplace-transf}
\begin{equation}
S(\tau) = \frac{S_0}{\left(1 + \frac{\tau}{\alpha}\right)^\alpha} = S_0 \frac{\alpha^\alpha}{(\alpha+\tau)^\alpha} \,. \label{S-tau}
\end{equation}
Since $s(x,\tau)=s_0(x)\exp(-x\tau)$, one obtains for the average susceptibility \eqref{av-susc}
\begin{equation}
\bar x(\tau) = \frac1{1 + \frac{\tau}{\alpha}} = \frac{\alpha}{1+\tau} \,, \label{bar-x-tau}
\end{equation}
which starts from $\bar x=1$ at the initial time $\tau=0$ and then decays to zero for increasing $\tau$,
slowing down the epidemic. Note that $s(x,\tau)$ obeys the scaling law \cite{Neipel:2020xxx}
\begin{equation}
s(x,\tau) = \bar x^{\alpha - 1} s_0(x/\bar x)\,.
\end{equation}
Thus, $s(x,\tau)$ is shape invariant, i.e., the gamma distribution is kept during the whole time evolution, 
instead of being just an initial condition.

\begin{figure}[H]
\centering
    \includegraphics[width=.5\linewidth]{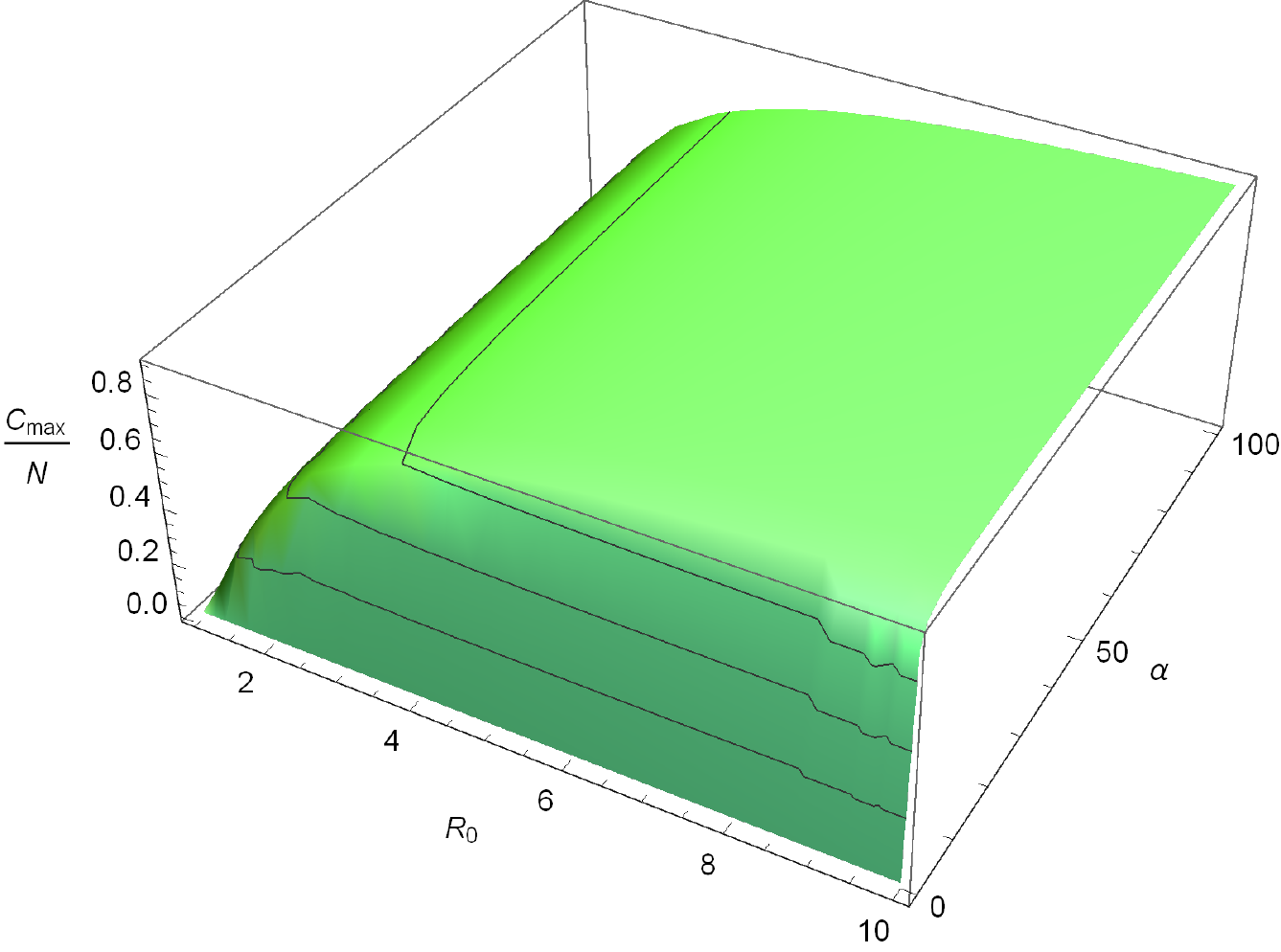}\hfill
    \includegraphics[width=.5\linewidth]{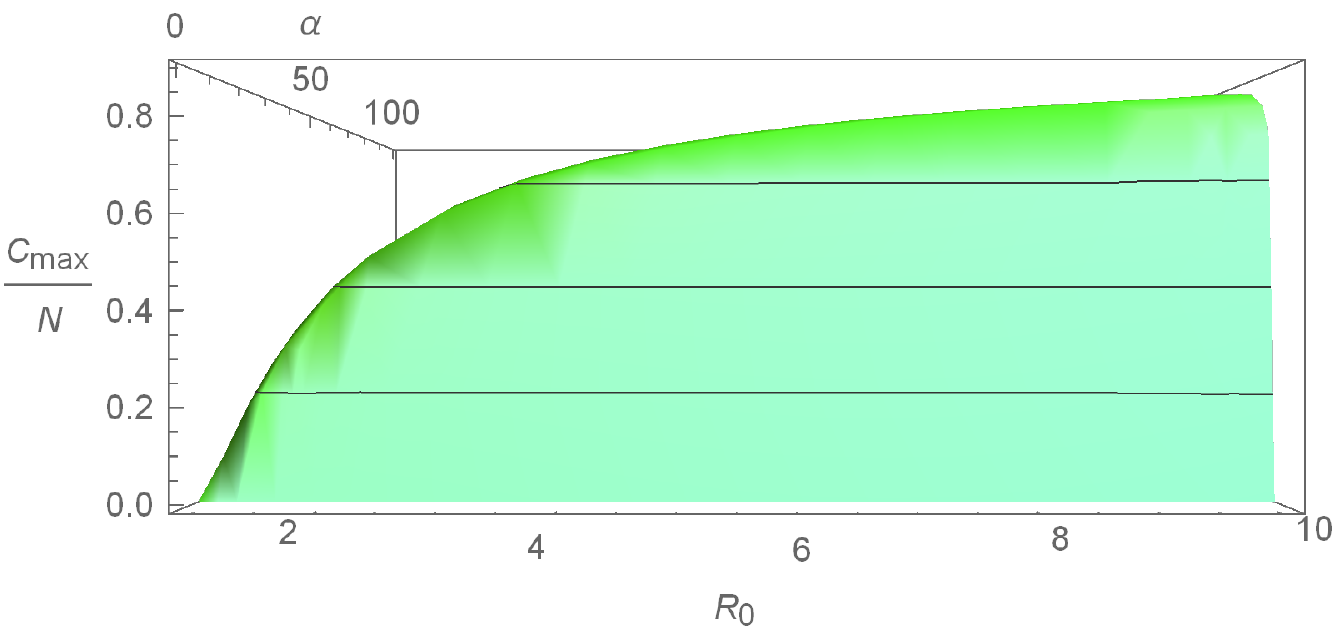}
    \caption{Three-dimensional plot of the herd immunity threshold $\frac{C_{\text{max}}}{N}$ of a heterogeneous population as a function of $R_0$ and $\alpha$ in the case of a one-parameter gamma distribution.}
    \label{fig:hva2}
\end{figure}

Using \eqref{S-tau} and \eqref{bar-x-tau}, equ.~\eqref{I(t)-hetero} can be rewritten as
\begin{equation}
\frac{dI}{d\tau} = S_0\left(1 + \frac{\tau}{\alpha}\right)^{-1-\alpha} - \frac{\gamma}{\beta}\,, \label{dotItau}
\end{equation}
which can be integrated to give
\begin{equation}
I(\tau) = I_0 + S_0\left[1 - \left(1 + \frac{\tau}{\alpha}\right)^{-\alpha}\right] - \frac{\gamma}{\beta}\tau\,. \label{Itau}
\end{equation}
This has a maximum for
\begin{equation}
\left(1 + \frac{\tau}{\alpha}\right)^{1+\alpha} = R_{\text e}\,,
\end{equation}
with $R_{\text e}$ the effective reproduction number defined in equ.~\eqref{ern}. The maximum
value of $I$ is given by
\begin{equation}
\frac{I_{\text{max}}}{S_0} = \frac{I_0}{S_0} + 1 - \frac1{R_{\text e}} - \frac1{R_{\text e}}(1 + \alpha)
\left(R_{\text e}^{\frac1{1+\alpha}} - 1\right)\,.
\end{equation}
In the special case $I_0\ll S_0\approx N$, this boils down to equ.~(18) of \cite{Neipel:2020xxx}. 
Then, the cumulative number of infections reads
\begin{equation}
C(\tau) = N - S(\tau) = N\left[1 - \left(1 + \frac{\tau}{\alpha}\right)^{-\alpha}\right]\,,
\end{equation}
which, evaluated at the maximum of $I$, becomes
\begin{equation}
\frac{C_{\text{max}}}N = 1 - R_0^{-\frac{\alpha}{1+\alpha}}\,. \label{herd-imm-hetero}
\end{equation}
This is the generalization of the herd immunity level \eqref{herd-imm-hom} to a heterogeneous population.
In Figure \ref{fig:hva2} we give a three-dimensional plot of the HIT as a function of $R_0$ and $\alpha$.

\vspace{0.5cm}

\begin{figure}[H]
    \centering
    \includegraphics[scale=1]{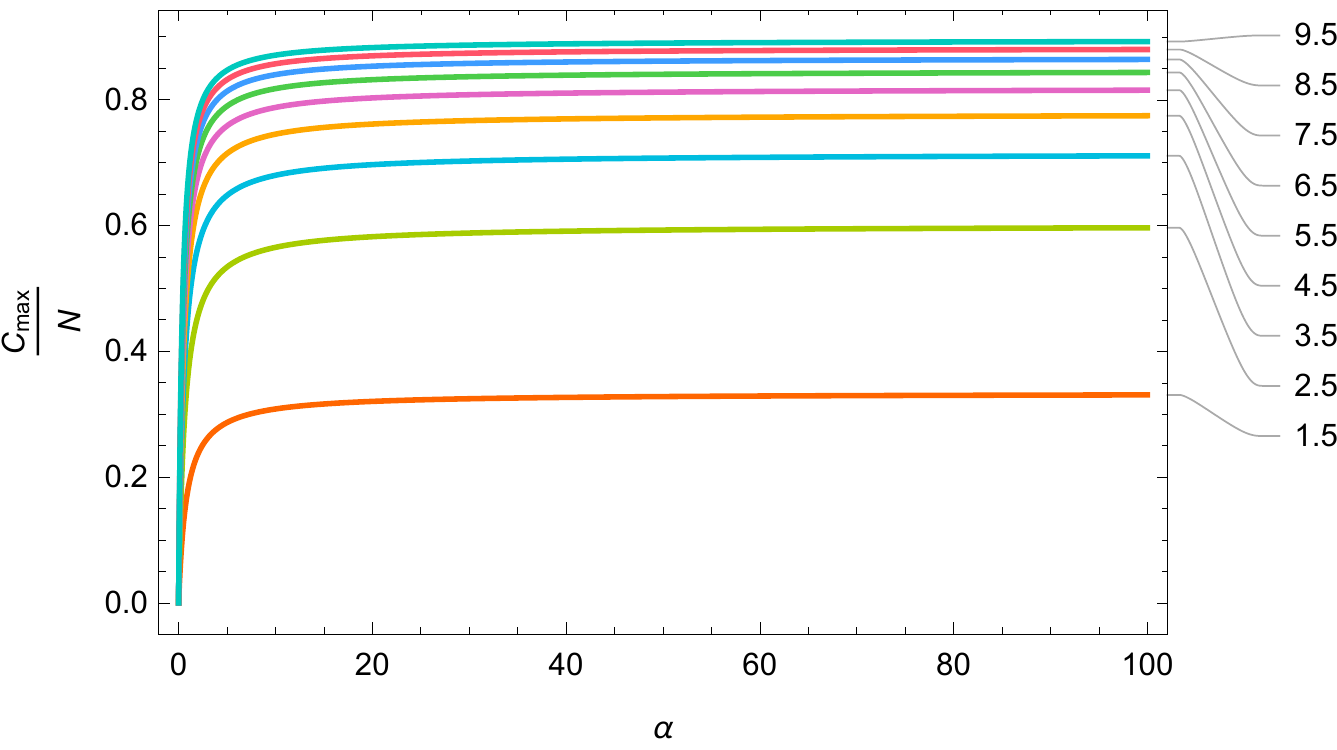}
    \caption{HIT $\frac{C_{\text{max}}}{N}$ of a heterogeneous population as a function of $\alpha$, for different values of $R_0$ (indicated on the right side of the plot), in the case of a one-parameter gamma distribution.}
    \label{fig:hva1}
    \vspace{1cm}
    \includegraphics[scale=1]{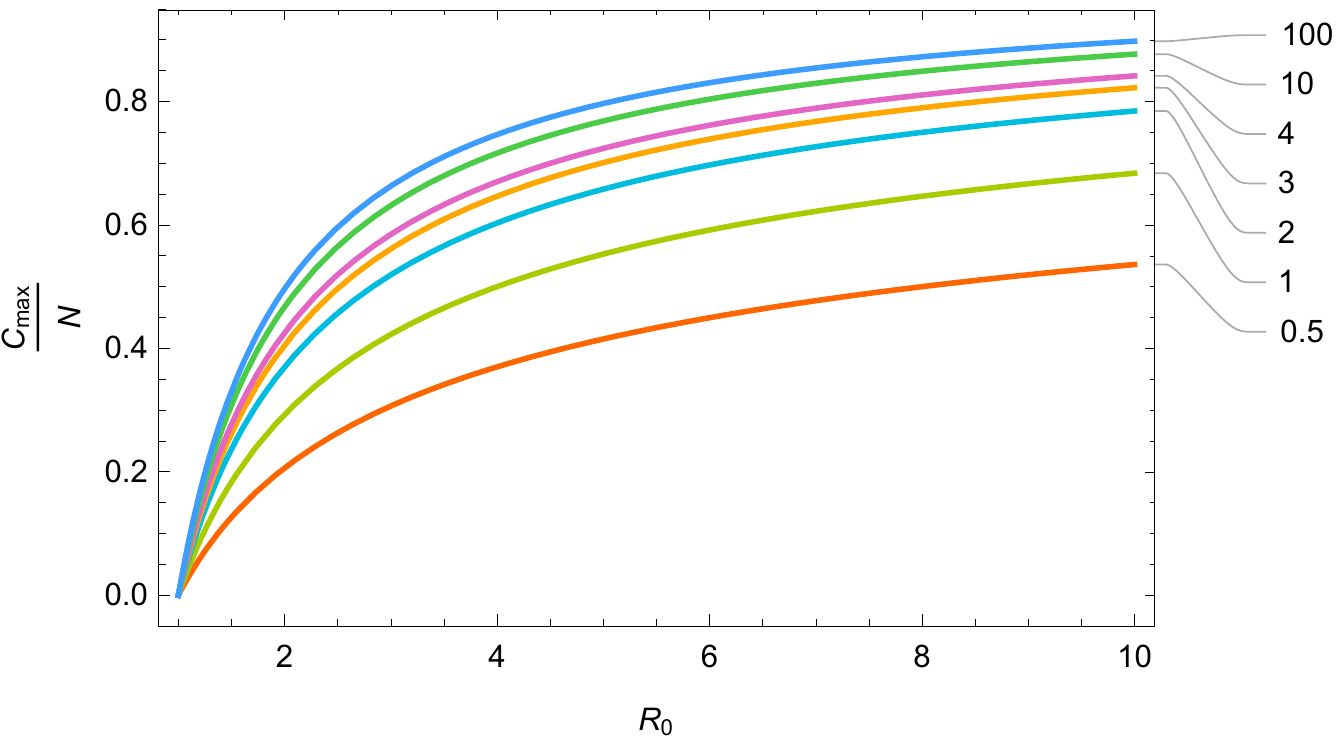}
    \caption{HIT $\frac{C_{\text{max}}}{N}$ of a heterogeneous population as a function of $R_0$, for different values of $\alpha$ (indicated on the right side of the plot), in the case of a one-parameter gamma distribution.}
    \label{fig:hvR0a}
\end{figure}

On the other hand, the plot in Figure \ref{fig:hva1} displays the herd immunity level \eqref{herd-imm-hetero} as a function of $\alpha$ for fixed values of $R_0$ (we consider $R_0=1.5, 2.5, 3.5, 4.5, 5.5, 6.5, 7.5, 8.5, 9.5$ as sampling values), while the plot in Figure \ref{fig:hvR0a} shows the HIT \eqref{herd-imm-hetero} as a function of $R_0$ for different values of $\alpha$ (we consider $\alpha=0.5, 1, 2, 3, 4, 10, 100$ as sampling values).

We observe that, for finite $\alpha$, \eqref{herd-imm-hetero} is smaller than \eqref{herd-imm-hom}, to which it
reduces in the homogeneous limit $\alpha\to\infty$.
In particular, for an $R_0$ between 2.5 and 4, herd immunity is expected to require 60-75 percent of the population in the homogeneous limit $\alpha\to\infty$, while these percentages drop to the range 26-50 percent for $\alpha$ between 0.5 and 1. In other words, for fixed values of $R_0$, the smaller $\alpha$, the more heterogeneous the population and the lower the threshold for achieving herd immunity.

\vspace{0.5cm}

\subsubsection{A possible way to take into account mitigation measures in model fitting}

When performing model fitting of $J$ (defined in equ. \eqref{Jeq}), one has to deal with data, let us call them $J_{\text{exp}}$, which are affected by mitigation measures (lockdown, etc.).
Here we propose a possible way to take this into account, which consists in fitting $J$ from the model (with the one-parameter initial gamma distribution), namely
\begin{equation}
J (\tau) = \beta I S_0 \left( 1 + \frac{\tau}{\alpha} \right)^{-\alpha -1} \,, 
\end{equation}
with $J_{\text{exp}}$ from data and solve a system of differential equations \textit{at the same time}, considering $\beta=\beta(t)$ (cf. \cite{Neipel:2020xxx}). 

We start from the differential equation for $\beta$, obtained by taking the derivative of $J$. Indeed, this provides a differential equations for $\beta$ (more precisely, for ${\rm{ln}} \beta$) if ${\rm{ln}}J(t)$ is given, which does not require knowledge of the amplitude of $J$. As $J(t)=J_{\text{exp}}$, we get
\begin{equation}\label{dotbeta}
\dot{\beta} = \beta \left[ \frac{d}{d t} {\rm{ln}} J_{\text{exp}} - \frac{\dot{I}}{I} + \left(\frac{\alpha+1}{\alpha} \right) \left(1 + \frac{\tau}{\alpha} \right)^{-1} \beta I \right] \,, 
\end{equation}
with
\begin{equation}
\frac{\dot{I}}{I} = \frac{dI}{d\tau} \frac{d\tau}{dt}\frac{1}{I} = \left[ S_0 \left( 1 + \frac{\tau}{\alpha} \right)^{-1-\alpha} - \frac{\gamma}{\beta} \right] \beta \,, 
\end{equation}
where we have used \eqref{dottaudef}
and \eqref{dotItau}.
Moreover, recall that, integrating the latter, we find \eqref{Itau}.
Plugging all of this back into \eqref{dotbeta}, we get the following differential equations:
\begin{equation}\label{diffeqsist}
\begin{split}
\dot{\beta} & = \beta \Bigg \lbrace \frac{d}{d t} {\rm{ln}} J_{\text{exp}} - \left[ S_0 \left( 1 + \frac{\tau}{\alpha} \right)^{-1-\alpha} - \frac{\gamma}{\beta} \right] \beta \\
& + \left(\frac{\alpha+1}{\alpha} \right) \left(1 + \frac{\tau}{\alpha} \right)^{-1} \beta \left[I_0 + S_0 \left( 1 - \left(1 + \frac{\tau}{\alpha} \right)^{-\alpha} \right) - \frac{\gamma}{\beta} \tau \right] \Bigg \rbrace \,, \\
\dot{\tau} & = \beta \left[I_0 + S_0 \left( 1 - \left(1 + \frac{\tau}{\alpha} \right)^{-\alpha} \right) - \frac{\gamma}{\beta} \tau \right] \,.
\end{split}
\end{equation}
These have to be solved while performing the fitting of $J$ from the model, that is
\begin{equation}\label{Jmodel}
J (\tau) = \beta S_0 \left[ I_0 + S_0 \left(1 - \left(1 + \frac{\tau}{\alpha} \right)^{-\alpha} \right) - \frac{\gamma}{\beta} \tau \right] \left( 1 + \frac{\tau}{\alpha} \right)^{-\alpha -1} \,,
\end{equation}
with the data $J(t)=J_{\text{exp}}$.

\subsection{HIT for the case of a two-parameter gamma distribution}

We will now generalize the discussion above to the case of the usual two-parameter gamma distribution. Therefore we assume that $s_0(x)$ is gamma-distributed with shape parameter $\alpha>0$ and rate parameter $\eta>0$,
\begin{equation}
s_0(x)= S_0\frac{\eta^\alpha}{\Gamma(\alpha)} x^{\alpha - 1} e^{-\eta x}\,.
\end{equation}
Note that the case previously discussed is obtained from the above setting $\eta=\alpha$.
For the Laplace transform \eqref{Laplace-transf} we get
\begin{equation}
S(\tau) = \frac{S_0}{\left(1 + \frac{\tau}{\eta}\right)^\alpha} = S_0 \frac{\eta^\alpha}{(\eta+\tau)^\alpha} \,. \label{S-tau2}
\end{equation}
Since $s(x,\tau)=s_0(x)\exp(-x\tau)$, now for the average susceptibility \eqref{av-susc} we obtain
\begin{equation}
\bar x(\tau) = \frac1{\frac{\eta}{\alpha} + \frac{\tau}{\alpha}} = \frac{\alpha}{\eta + \tau}\,. \label{bar-x-tau2}
\end{equation}
Note that for $\tau=0$ we have $\bar{x}(0)=\alpha/\eta$. Then, using \eqref{S-tau2} and \eqref{bar-x-tau2}, equ.~\eqref{I(t)-hetero} can be rewritten as
\begin{equation}
\frac{dI}{d\tau} = S_0 \alpha \eta^\alpha (\eta+\tau)^{-1-\alpha} - \frac{\gamma}{\beta}\,, \label{dotItau2}
\end{equation}
which, integrated, yields
\begin{equation}
I(\tau) = I_0 + S_0\left[-\frac{\eta^\alpha}{(\eta+\tau)^\alpha}\right] - \frac{\gamma}{\beta}\tau\,. \label{Itau2}
\end{equation}
The latter has a maximum for
\begin{equation}
\eta^{-\alpha} \frac{(\eta+\tau)^{1+\alpha}}{\alpha} = R_{\text e}\,.
\end{equation}
Consequently, considering $S_0\approx N$, the cumulative number of infections reads
\begin{equation}
C(\tau) = N - S(\tau) = N\left[1 - \frac{\eta^\alpha}{(\eta+\tau)^\alpha}\right]\,,
\end{equation}
which, evaluated at the maximum of $I$, becomes
\begin{equation}
\frac{C_{\text{max}}}N = 1 - \eta^\alpha \left(\alpha \eta^\alpha R_0 \right)^\frac{-\alpha}{1+\alpha} = 1 - \left( \frac{\alpha}{\eta} R_0 \right)^\frac{-\alpha}{1+\alpha} \,. \label{herd-imm-hetero2}
\end{equation}
This is the generalization of the herd immunity threshold \eqref{herd-imm-hom} to the case of a heterogeneous population with an initial susceptibility given by a two-parameter gamma distributions. In the limit $\alpha/\eta \rightarrow 1$ we recover the HIT of the particular case of the initial one-parameter gamma distribution \eqref{herd-imm-hetero}.
The HIT homogeneous for a homogeneous population \eqref{herd-imm-hom} is recovered in the limit $\alpha/\eta \rightarrow 1$, $\alpha \rightarrow \infty$.
Besides, we can observe that, in case of the initial two-parameter gamma distribution case, a redefinition of $R_0$ emerges. Indeed, \eqref{herd-imm-hetero2} may be rewritten as
\begin{equation}
\frac{C_{\text{max}}}N = 1 - \hat{R}_0^\frac{-\alpha}{1+\alpha} \,, \label{herd-imm-hetero22}
\end{equation}
where we have introduced
\begin{equation}
    \hat{R}_0 = \hat{R}_0 (R_0,\alpha,\eta) := \frac{\alpha}{\eta} R_0 \,. \label{hatR0}
\end{equation}

\begin{figure}[H]
    \centering
    \includegraphics[width=.5\linewidth]{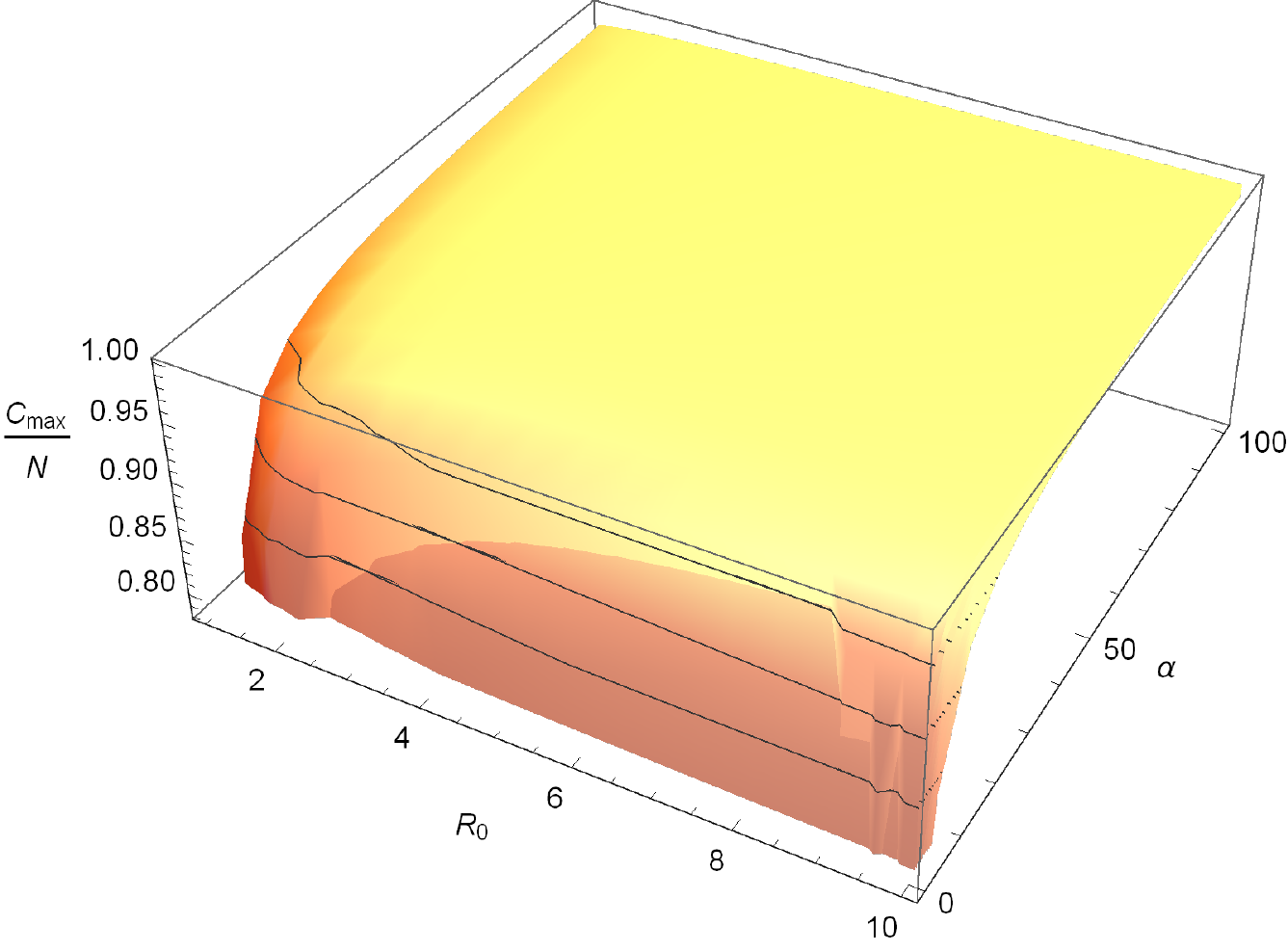}\hfill
    \includegraphics[width=.5\linewidth]{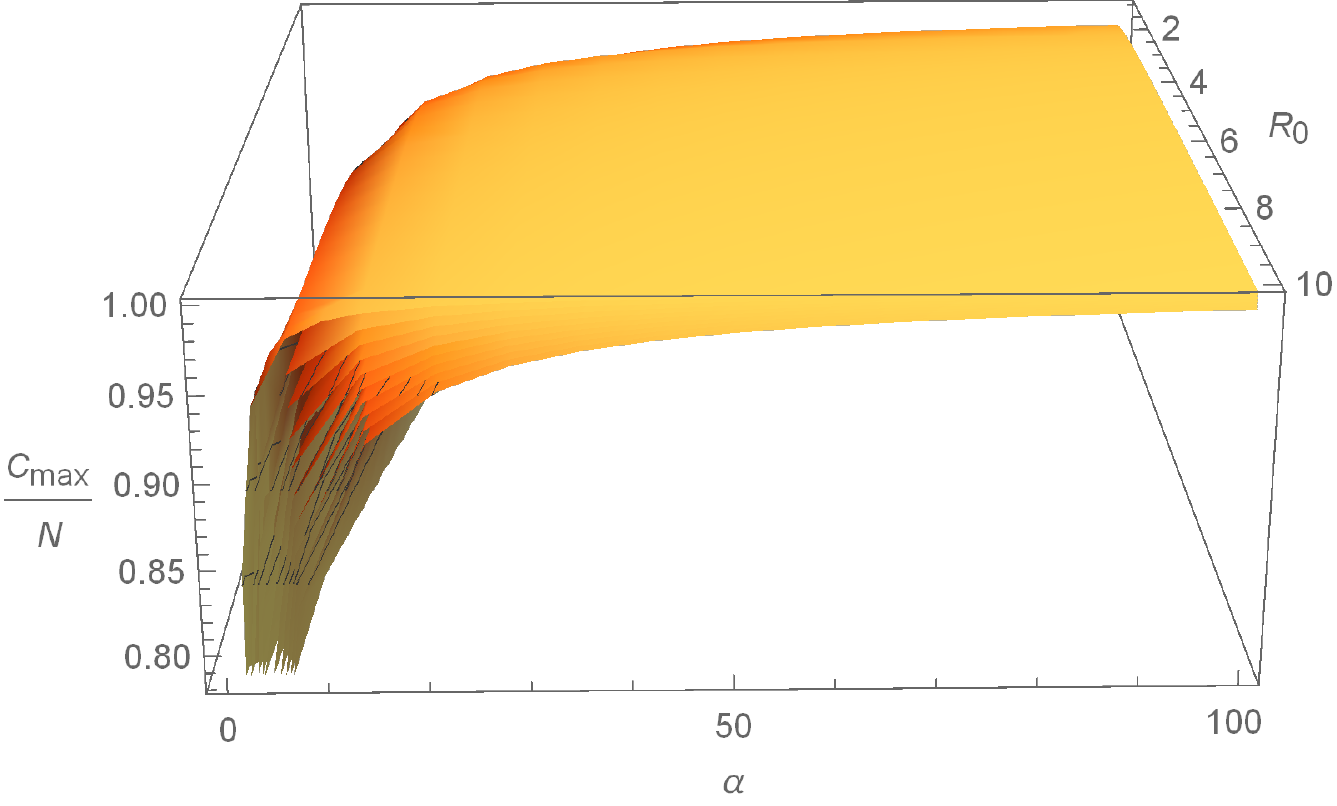}
    \caption{Three-dimensional plot of the HIT $\frac{C_{\text{max}}}{N}$ of a heterogeneous population as a function of $R_0$ and $\alpha$ in the case of a two-parameter gamma distribution, for $1 \leq \eta \leq 10$.}
    \label{fig:hvae}
    \vspace{1cm}
    \includegraphics[scale=1]{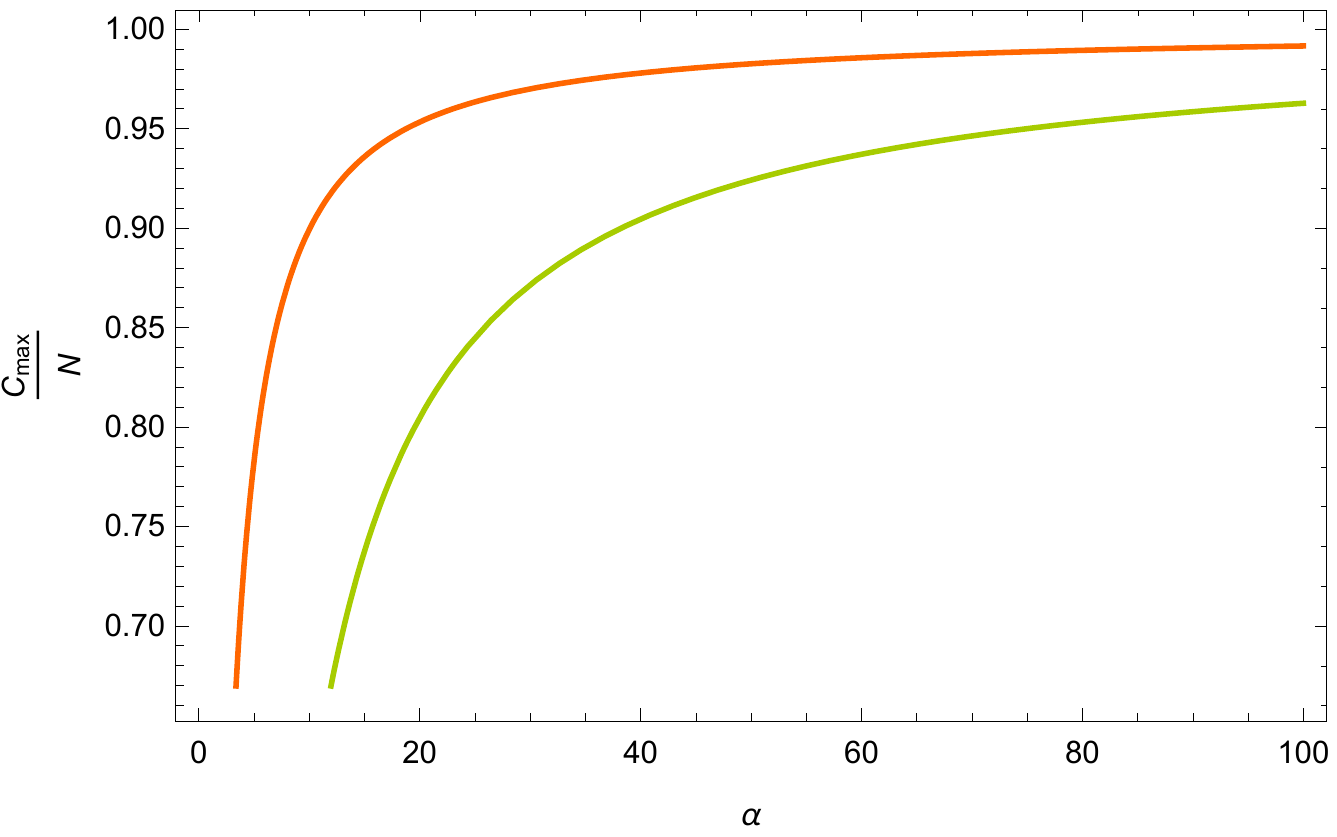}
    \caption{HIT $\frac{C_{\text{max}}}{N}$ of a heterogeneous population, for $R_0=2.5$, as a function of $\alpha$ in the case of a two-parameter gamma distribution, for two different values of $\eta$. The orange curve originates from taking $\eta=2$, while the green one from $\eta=9$.}
    \label{fig:hvae2}
\end{figure}

In Figure \ref{fig:hvae} we give a three-dimensional plot of the HIT as a function of $R_0$ and $\alpha$ as $\eta$ runs from 1 to ten. We can see that this outlines different surfaces depending on the value of $\eta$. The larger $\eta$, the higher the threshold is raised to achieve herd immunity, with the same $R_0$ and $\alpha$.
This is better highlighted in Figure \ref{fig:hvae2}, where we plot the herd immunity level $\frac{C_{\text{max}}}{N}$ of a heterogeneous population, with $R_0=2.5$, as a function of $\alpha$ for two different values of $\eta$ (we take as sample values $\eta=2$ and $\eta=9$).

\vspace{0.5cm}

Of course, as it is commonly understood also for $R_0$, for the $\hat{R}_0=\hat{R}_0 (R_0,\alpha,\eta)$ defined in \eqref{hatR0} we have that if $\hat{R}_0 <1$, each existing infection causes less than one new infection. In this case, the disease will decline and eventually die out. If $\hat{R}_0=1$, each existing infection causes one new infection. The disease will stay alive and stable, but there will not be an outbreak or an epidemic. On the other hand, if $\hat{R}_0>1$, each existing infection causes more than one new infection. The disease will be transmitted between people, and there may be an outbreak or epidemic.

For the sake of completeness, let us also say that the maximum value of $I$ in this case is
\begin{equation}
    \frac{I}{S_0} = \frac{I_0}{S_0} + \frac{\eta}{R_e} - \frac{1}{R_e} \left[ \left(\alpha \eta^\alpha R_e \right)^\frac{1}{1+\alpha} - \eta^\alpha R_e  \left(\alpha \eta^\alpha R_e \right)^\frac{-\alpha}{1+\alpha} \right] \,.
\end{equation}
All of this suggests that the parameter $\eta$ may take into account mitigation measures, which in fact affects both the average susceptibility (in particular, the initial one) and the HIT. Such mitigation measures can be even natural, that is driven by the epidemic history of the human being. For fixed $R_0$ and $\alpha$, decreasing $\eta$, which can also be understood as increasing the scale parameter $\theta$ of the initial gamma distribution, can be translated into contrasting heterogeneity effects, rising the HIT, and vice versa.
For fixed $R_0$, if $\eta$ is bigger enough with respect to $\alpha$, its presence lowers the herd immunity threshold.
For instance, in the two-parameter initial gamma distribution case, for an $R_0$ between 2.5 and 4 and values of $\alpha$ between 0.5 and 1, herd immunity is expected to require 7-29 percent of the population for $\eta$ between 1 and 2, with respect to the range 60-75 percent of a homogeneous population, with the same $R_0$, and the range 26-50 percent for a heterogeneous population with $\eta=\alpha$ between 0.5 and 1.

\section{Final remarks}\label{final}

In this paper we have elaborated on SIR-type models of epidemics, especially in regards to population heterogeneity effects, which capture the fact that the probability of being infected with a transmissible disease is not the same for all individuals of a population.

We have first reviewed homogeneous SIR-models, presenting an intriguing correspondence between the epidemic trajectory in a homogeneous SIR-type model and radial null geodesics in the Schwarzschild spacetime. The correspondence between the SIR model and radial null geodesics in the Schwarzschild geometry may be useful to obtain new analytically tractable epidemiological models,
e.g. by considering timelike and/or nonradial geodesics. Moreover, one could construct an action
principle that gives rise to the eqns. \eqref{SIR-common}.

Subsequently, we have analyzed modeling of population heterogeneity effects. We have first reviewed and discussed the case in which the initial susceptibility is given in terms of a one-parameter gamma distribution, deriving the associated HIT. The study has been done without considering mitigation measures and reinfections, which means, in particular, that the threshold to reach herd immunity has been estimated by considering natural infections without restrictions on individuals (lockdown, social distancing, etc.) and without taking into account possible vaccinations. 
Consequently, in the same setup we have also described a possible way to take into account mitigation measures when performing model fitting. Our proposal consists in considering a time-dependent infection rate $\beta=\beta(t)$ to derive a differential equation for it (actually, for ${\rm{ln}}\beta$) to be solved while implementing the fitting of $J$ from the model with $J_{\text{exp}}$ from data ($J$ is the number of new cases per day). 

Finally, we have derived the HIT in the case of an initial two-parameter gamma distribution. We find that the additional parameter ($\eta$) induces a shift of $R_0$, which can be interpreted as a possible way of taking into account mitigation measures in this setup. One could also say that $\eta$ allows to take into account the effects of whatever makes a population more or less heterogeneous.

To conclude, although the gamma distribution has been shown to be better than other ones in developing epidemiology models, it would be interesting to try to compute analytically the HIT for distributions other than gamma, in the context of heterogeneous SIR-type models. Consequently, an fundamental study would be comparison with data.
Besides, a critically important question is: how variable are humans in their susceptibility and exposure to transmissible diseases (such as SARS-CoV-2)? Hitherto, there is no definite answer to this question. Such issue, which, in particular, can be translated, in the context of this paper, to determine the value of $\alpha$ from model fitting, is left to a future investigation on heterogeneous models of epidemic.

\section*{Acknowledgements}

We thank F. Lingua for stimulating discussions. This work was supported partly by INFN and by MIUR-PRIN contract 2017CC72MK003.
L.~R.~would like to thank the Department of Applied Science and Technology of Politecnico di Torino and the INFN for financial support.

\end{document}